\documentclass[12pt]{iopart}
  \usepackage{srcltx}
  \usepackage{iopams}
  \usepackage{t1enc}
  \usepackage[latin1]{inputenc}
  \usepackage{verbatim}

  \usepackage[dvips]{graphicx,color} 
  \usepackage{float, times, amsbsy}
  \usepackage{amssymb}
  \usepackage{calc}

\begin{document}


\title[Proof of concept]{Cooling by heating a nanodroplet - proof of concept}

\author{J. J. Papini$^1$}
\address{DNRF Centre ``Glass and Time'', IMFUFA, Dept. of Sciences, Roskilde University, P.O. Box
260, DK-4000, Denmark.}
\ead{$^1$ papini@ruc.dk}


\begin{abstract}

\noindent Recently \cite{papini_1} predicted the existence of an intriguing new phenomenon. It was
shown that if temperature is suddenly raised at the surface of a sphere the temperature in the
interior initially decreases. The authors of \cite{papini_1} gave a thorough analysis explaining the
physics leading to this remarkable effect. They linked the existence of the phenomenon to the subtle
thermomechanical coupling between displacement and temperature in a sphere that is able to expand
freely and showed that the effect is expected to be largest close to the glasstransition temperature
of the perturbed material. The prediction was based on the assumption of quasi-elasticity where the
sample is much smaller than the wavelength of acoustic waves at frequencies of relevance. Being in
the inertia free limit they could ignore the acceleration term in the thermoviscoelastic equations
of motion. Here we give the first empirical proof of existence of the effect by performing molecular
dynamics simulations of a supercooled Kob-Andersen binary mixture of Lennard-Jones particles forming
a droplet consisting of 500.000 particles. We show that the phenomenon is real even when inertia
cannot be disregarded. 


\end{abstract}
\maketitle



\section{Introduction}
\noindent The thermoviscoelastic nature of viscous liquids gives rise to a rich phenomenology
\cite{harrison} and pose fundamental challenges in science \cite{dyre1}. Recently
\cite{Tage_1,Tage_2} discuss the breaking of hydrostatic conditions that occurs when the release of
stresses that arise from thermal waves take place on the timescale of experiment (the stress tensor
is temporarily nonisotropic). Papini and coauthors \cite{papini_1} analyze this breaking of
hydrostatic conditions and its consequences pertaining to two cases of heating a spherical sample
through the mechanically free surface. E.g. as a finite amount of heat is added at the surface, the
interior of the sphere, even close to the surface, instantaneously cools to a common temperature
independent of radius; the expansion of the surface is immediately felt in the interior and as no
heat has been transfered it cools adiabatically. They explain the phenomenon and give the conditions
for its existence in terms of a subtle thermomechanical coupling, that peaks in the glass transition
region of a material.\\

\noindent It all boils down to the nature of longitudinal expansion, which is present when you heat
a sphere homogenously at its surface. The thermomechanical coupling can be quantified by the
relative difference between the longitudinal specific heat, $c_l$ and the isobaric specific heat,
$c_p$, that are defined in Eq's (\ref{defcl}) and (\ref{defcp}):

\begin{equation}
 c_l=\frac{M_S}{M_T} c_V\label{defcl}
\end{equation}
\begin{equation}
 c_p=\frac{K_S}{K_T} c_V\label{defcp},
\end{equation}

\noindent where $K_S$ and $K_T$ are the adiabatic and isothermal bulkmoduli respectively and $c_V$
is the isochoric specific heat. The longitudinal moduli are given by $M_S=K_S+4/3 \, G$ and
$M_T=K_T+4/3 \, G$ with $G$ being the shear modulus. The authors of \cite{papini_1} point out that
the difference between $c_l$ and $c_p$ only occurs when two conditions are met simultaneously.
First, shearmodulus must be nonvanishing compared to bulkmodulus and second, the adiabatic and
isothermal bulkmoduli must differ significantly. This situation will be most probable near the glass
transition where the constitutive quantities become frequency dependent. Combining Eq's.
(\ref{defcl}) and (\ref{defcp}) they show that the effect they term ``Cooling by heating'' only
occurs when the resulting thermo-mechanical coupling constant $a$ defined as

\begin{equation}
a\equiv \frac{c_p-c_l}{c_p}
=\frac{4}{3}\cdot \frac{G}{M_T}\cdot \frac{c_p-c_V}{c_p}\label{specheatdiff},
\end{equation}

\noindent differs from zero.\\

\noindent The analyzis given in \cite{papini_1} is based on the assumption of quasi-elasticity where
they could ignore the acceleration term in the thermoviscoelastic equations of motion for
temperature $\delta T(r,s)$ and displacement $\textbf{u}(r,s)=u(r,s)\textbf{r}$ in the frequency
domain \cite{Tage_2}:

\begin{equation}\label{momentumbalance}
\boldsymbol{\nabla}\left( M_T\boldsymbol{\nabla} \cdot \textbf{u}-  \alpha_p K_T \delta T
\right)=\rho s^2\textbf{u}
\end{equation}
\begin{equation}\label{energybalance}
c_Vs\delta T+T_0 \alpha_p K_T \boldsymbol{\nabla} \cdot \textbf{u} = \lambda
\boldsymbol{\nabla}^2\delta T
\end{equation}

\noindent Here $\alpha_p\equiv \frac{1}{V} \left(\frac{\partial V}{\partial T} \right)_p$ is the
isobaric expansion coefficient, $\lambda$ the heat conductivity as defined by Fourier's law,
$s=i\omega$ the Laplace frequency with the cyclic frequency $\omega$ and $\rho$ the average mass
density.\\

\noindent Introducing the potential function $\phi$ by $\textbf{u}=\boldsymbol{\nabla} \phi$ Eq.
(\ref{momentumbalance}) becomes 

\begin{equation}\label{momentumbalance2}
\boldsymbol{\nabla}\left( M_T\boldsymbol{\nabla}^2 \phi- \rho s^2 \phi- \alpha_p K_T \delta T
\right)=0.
\end{equation}

\noindent Adjusting $\phi$ with an appropriate (r-independent) additive constant, the integration
constant that comes out integrating Eq. (\ref{momentumbalance2}) is set to zero. Introducing the
dimensionless quantities $\alpha=T_0\alpha_p$, $c=T_0c_l/K_T$, $g=4G/3K_T$,
$1/\gamma_l=M_T/M_S=c_V/c_l=1-\alpha^2/c(g+1)$ as well as the acoustic (adiabatic) and thermal wave
vectors defined by $q^2=\rho/M_Ss^2$ and $k^2=c_ls/\lambda$ respectively, Eq.'s
(\ref{momentumbalance2}) becomes, on dimensionless form:

\begin{equation}\label{momentumbalancedim}
\left(\boldsymbol{\nabla}^2 -\gamma_l\left(\frac{q}{k} \right)^2 
\right)\phi=\frac{\alpha}{1+g}\delta T
\end{equation}

\noindent As long as $q^2/k^2\ll 1$ we can safly assume non-inertial conditions. Introducing
characteristic numbers for the speed of sound $c_{sound}=M_S/\rho\approx 10^3 m/s$ and diffusivity
$D=c_l/\lambda\approx10^{-7} m^2/sec$ we have
$\left(\frac{q}{k}\right)^2=\frac{D}{c_{sound}^{-2}}s=2\pi 10^{-13}\sec f$, $f$ being the frequency.
When the system studied becomes extremely small, like in a nano-sized droplet, the relevant
timescales (picoseconds) are exactly those where the corresponding frequencies give a
$\left(\frac{q}{k}\right)^2\approx1$. These frequencies, which give wavelengths,
$\lambda=c_{sound}/f$, comparable to the size of the sample (nanometers) are also those that make
the assumption of adiabatic propagation of acoustic waves flawed. Therefore the situation becomes
much more complicated when dealing with very small samples than in the quasi-elastic case studied in
\cite{papini_1}.\\

\noindent To see if the phenomenon predicted in \cite{papini_1} is also present in nano-sized
systems we performed molecular dynamics simulations of a small droplet (described below) thus giving
a proof of concept regarding the existence of Cooling by heating.

\section{Simulations of a nano-sized droplet}

\subsubsection{Preparing the simulations}
\noindent The cooling by heating in a droplet is expected to be biggest at the glass transition
in a supercooled fluid.  Almost all Molecular Dynamics (MD)  simulations \cite{toxmd} of supercooled
systems are performed for the Kob-Andersen binary mixture of Lennard-Jones particles (KABLJ)
\cite{KA}. This system consist of
80\% A-particles and 20\% smaller B-particles in a strong exothermic mixture, which makes it
very resistant against crystallization \cite{toxmix}. The glas transition temperature, $T_g$ is
 estimated to be $T_g$=0.438 \cite{KA} for an uniform system of
$N=N_{\textrm{A}}+N_{\textrm{B}}$=1000
Lennard-Jones (LJ) particles at a density, $\rho=1.2$. A droplet of only one thousand particles
is, however, much too small to establish the effect and to measure it
accurately, and for this reason we have created a droplet of $N$=500 000  particles.
The reason for choosing such
 a big number is to make sure that there is time enough to measure an
 effects of thermomechanical coupling before everything is washed out by the
 diffusion of heat.

   A  droplet of $N$=500 000 KABLJ particles was created in the following way:
 A  KABLJ system of $N$=1000 particles   was calibrated
 at a density of $\rho=1.2$ and temperature $T$=0.40. A number of copies
 of the system were then fused into a huge system, from which a spherical droplet
 consisting of 500.000 particles was cut out.
 The droplet was then placed in a  container with the
 box-side length 10 times bigger than the radius of the droplet \cite{toxbox}. Droplets
 were then equilibrated at 3 temperatures: $T$=0.38,0.40,0.42, all below the glas-temperature.
In figure
 \ref{fig.1a} we show the density profile of a droplet at the temperature $T=0.40$ with
 the density of the bulk $\rho=1.08$. We took the relaxation time,
 $\tau_{\alpha}$ of the systems to be given by the time when the AA incoherent
 intermediate scattering function in a KABLJ
  is equal to $1/e$ (at wave vector $q=7.25(\rho/1.2)^{1/3}$ \cite{ulp}).
 Then we run simulations at the four state points, dumping configurations separated
 by $2\tau_{\alpha}$, creating an ensemble consisting of 300 droplets with uncorrelated  start
configurations at each state point.

\subsubsection{Production runs}
\begin{figure}[H]
\begin{center}
\includegraphics[scale=0.4]{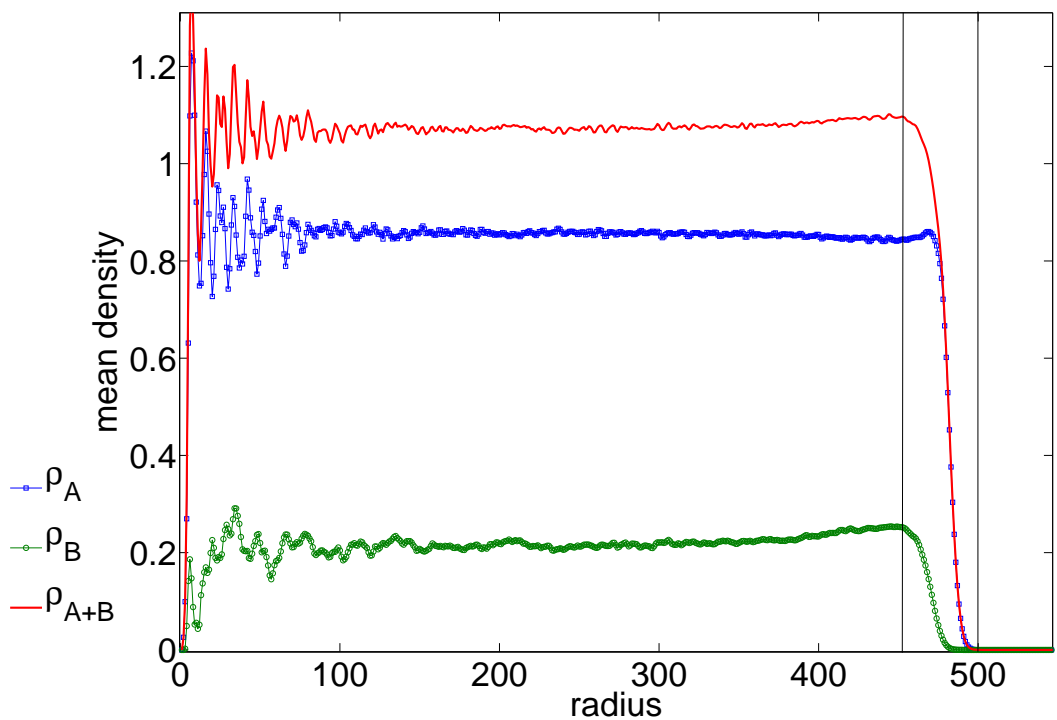}
\caption{Density profiles, $\rho(r)$, as a function of the distance $r$ from the center
  in the droplet before the thermal wave was created. The KABLJ droplet consists of  solvent
A-particles and solule B-particles.
The droplet were heated up in the surface shell at $r \approx 500 \sigma_{\textrm{AA}}$, marked by
the vertical lines }\label{fig.1a}
\end{center}
\end{figure}

\noindent   In the production runs we thermostatted the particles in the surface shell  between
concentric spheres at the surface of the droplet,
  keeping track of all the particles during the whole run. The particles
 in the  shell  were defined as those that are situated in an  interval of radii  from the center
 shown by  vertical lines in figure \ref{fig.1a}.
These particles were thermostated by a Nose-Hoover thermostat (NHT) \cite{toxmd}.
The NHT thermostat consists of a "friction therm" in the classical  equations of motion
and the thermostat has a relaxation parameter $\tau_T$ which controls the heat flow into and out of
the system. A big value of
 $\tau_T$
results in big and slow oscillations of the temperature $T$ around the target temperature whereas
a small value of  $\tau_T$  results in small and quick oscillations  \cite{explain}. The NHT
  thermostated particles  are not  accelerated (heated) or damped (cooled) by
particle collisions, but  by the
dynamic  friction parameter in the equations of motion,
 by which one in general avoids thermal waves in the system if all particles are coupled to the
thermostate.
In the present simulations we want, however, to create a thermal wave by heating only the particles
at the surface.

\begin{figure}[H]
\begin{center}
\includegraphics[scale=0.4]{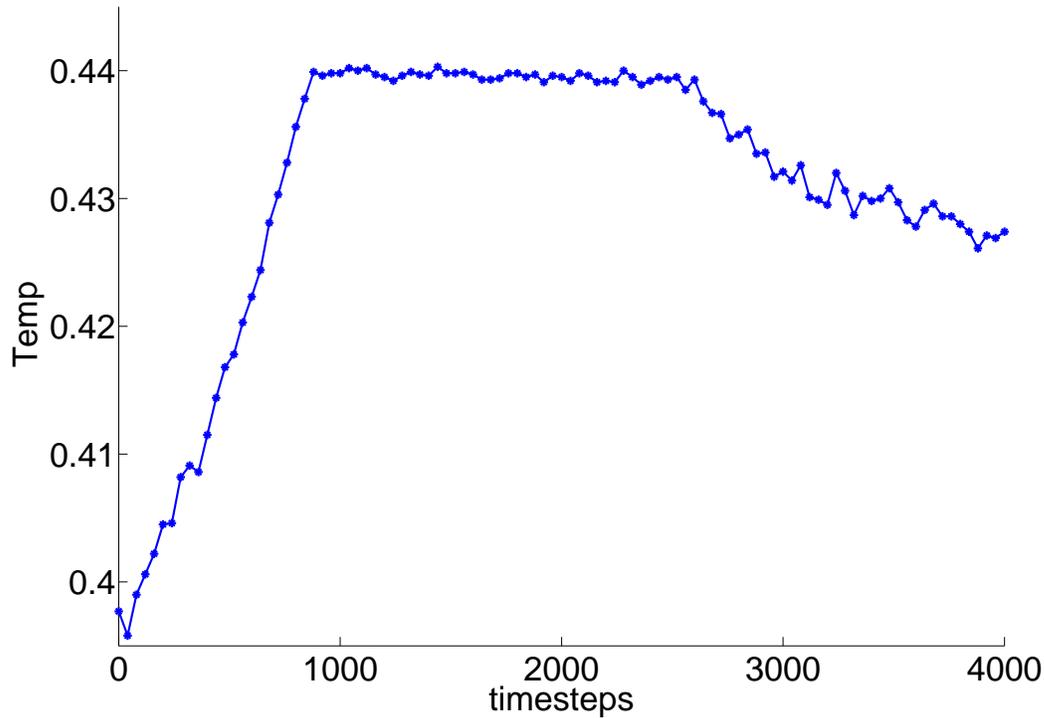}
\caption{ Temperature evolution in the thermostated surface shell at the start of the thermal wave
into the droplet.
After a short heating of $\approx$ 1000 timesteps with a big value of $\tau_T$, the shell was
calibrated 2000
timesteps wit $\tau_T \approx \tau_c$. from where the  surface thermostate was swiched
off.}\label{fig.1b}
\end{center}
\end{figure}

The thermal wave was  started by increasing the thermostate temperature with
   $\delta T=0.02$,  for particles in the surface shell.
 This was done in the following way. First we use  a big value of $\tau_T$ for a short time
 by which the temperature change were ``Heaviside-like''. At the time
 where the temperature reached the desired value, we switched to a smaller value of $\tau_T$ which
 immediately stabilized the temperature at the new  target temperature ($T+\delta T$) in the surface
shell.
 The NHT thermostating with the small value of $\tau_T$ were then continued
  for a short time in order
 to pump some heat into the surface shell. Then  the thermostat were switched of, giving a constant
 energy in the droplet for  the rest of the simulation and at the same time starting a heat wave
into the droplet. The temperature evolution in
the surface shell  is shown in figure \ref{fig.1b} for the step done
 from $T=0.40$ to $T=0.44$.\

\noindent In order to calculate the local quantities, like density and temperature we divided
 the system into concentric shells, with some given thickness. In all the figures shown in
 this work we chose to divide the droplets into 10 shells. This number is based on a compromise
 to secure good enough statistics, especially in  the innermost shells, but still get a good
 enough resolution in order to study the evolution of the temperature  in the shells.


\section{Results from simulations of a nanosized droplet - proof of existence}

\noindent  The evolution of temperature and displacement fields
 were  analyzed in  \cite{papini_1}. The analyse included  the deduced evolution of stresses and
pressure after a spherical
 system has been perturbed thermally at its surface.  Only addition of heat, or jumps up in
temperature at the surface of the sphere
were, however, considered. They do not consider what effect a removal of heat, or jump down in
temperature will affect.
 In the thermoelastic case (solid) where one can ignore memory effects and the influence of
temperaure on relaxation properties
 of the system, one would expect the phenomenon to be symmetric with respect to the
 direction of heat transfer. In the case of a system whose relaxation properties change with
temperature, like in a super cooled liquid, one can, however,
 not apriori
 expect a symmetric behavior because the relaxation time
varies exponentially with thetemperature.

\begin{figure}[H]
\begin{center}
\includegraphics[scale=0.6]{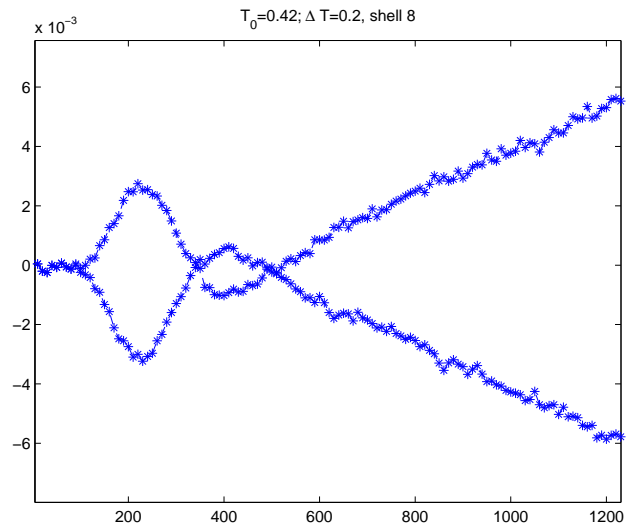}
\caption{\label{fig.2a1} Temperature evolution in an inner shell (No. 8) in the droplet after the
surface shell has beeen heated/cooled. The effect is symmetric with respect of the sign of the
temperature step at the surface}
\end{center}
\end{figure}

 In figure \ref{fig.2a1} we show the temperature of a
section of the droplet close to
 the surface. Here we first see the  wave induced by heating or cooling at the surface followed
first by
 the  reverse characteristic cooling/heating phenomenon which again is followed by the diffusion of
heat. But the phenomenon is symmetric despite the
exponentially varying relaxation times in the supercooled rigime.
 

\begin{figure}[H]
\begin{center}
\includegraphics[scale=0.4]{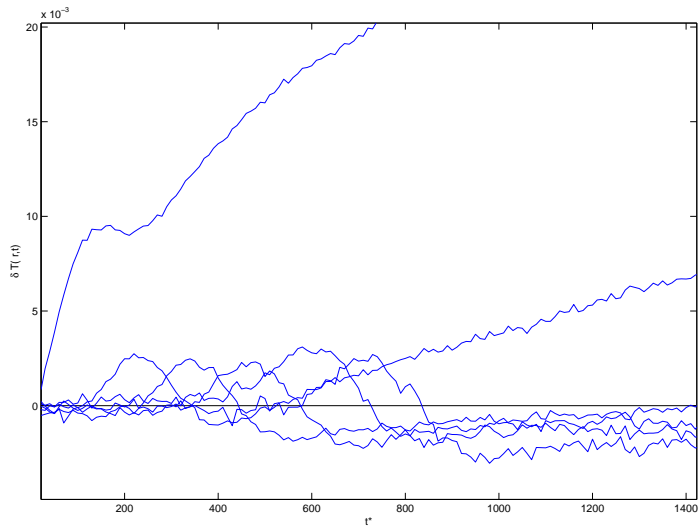}
\caption{\label{fig.2b} The temperature change as a function of time for a number of radii. The
uppermost curve represents the evolution in temperature close to the surface. The curve below refers
to a a radius a bit closer to the center of the droplet, and so on. For all radii one first sees the
elastic wave that is induced by the step in temperature on the surface of the droplet, followed by
the an increasing temperature due to the diffusion of heat. However, for the innermost parts of the
droplet the elastic wave is followed by the phenomenon of Cooling by heating. The
size of the effect increases the closer to the center you look.}
\end{center}
\end{figure}

\noindent Figure \ref{fig.2b} shows the temperature as a function of time for a number of concentric
 shells in the droplet. Whe the temperature at the surface  is increased  it starts an acoustic wave
which
 travels in towards the center of the sphere and results in an increase in temperature due to
 the thermomechanical coupling. This is followed by a temperature drop - cooling by heating,
 thus providing a proof of the concept introduced in \cite{papini_1}. We have left out the innermost
 shells, i.e. shells 1 and 2. The reason for this is twofold. First of all, since the number of
 particles in the shells - they all have the same thickness, decreases as one gets closer to the
 center of the droplet, the noise increases as there are fewer particles to average over.
 The second reason is that in the innermost parts of the droplet, the acoustic and thermal waves
mix, giving a more messy picture.

\section{Discussion}


\ack

U. R. Pedersen is greatfully acknowledged for sharing his ideas on how to approach the simulations.
Center for viscous liquid dynamics ``Glass and Time'' is sponsored by The Danish National Research
Foundation.



\begin{thebibliography}{3}
\bibitem{explain} A natural time unit for a  MD thermostate is the mean collision time $\tau_c$
 of the thermostated particles. A small value of the thermostates relaxation time, $\tau_T$,
corresponds to  $\tau_T \approx \tau_c$.
\end{thebibliography}
\end{document}